\documentclass[twocolumn,showpacs,preprintnumbers,amsmath,amssymb,prl]{revtex4}

\usepackage{graphicx}
\usepackage{bm}

\begin{document}

\preprint{}

\title{Scaling of degree correlations and the influence on diffusion in scale-free networks}

\author{Lazaros K. Gallos}
\author{Chaoming Song}
\author{Hern\'an A. Makse}
\affiliation{
Levich Institute and Physics Department,
City College of New York, New York, NY 10031, US}

\date{\today}

\begin{abstract}
Connectivity correlations play an important role in the structure of
scale-free networks.  While several empirical studies exist, there is
no general theoretical analysis that can explain the largely varying
behavior of real networks. Here, we use scaling theory to quantify the
degree of correlations in the particular case of networks with a
power-law degree distribution.  These networks are classified in terms
of their correlation properties, revealing additional information on
their structure. For instance, the studied social networks and the
Internet at the router level are clustered around the line of random
networks, implying a strongly connected core of hubs. On the contrary,
some biological networks and the WWW exhibit strong anti-correlations.  The
present approach can be used to study robustness or diffusion, where
we find that anti-correlations tend to accelerate the diffusion
process.

\end{abstract}


\maketitle

The topological structure of complex networks is largely determined by
the way in which the constituent units are interconnected.
Correlations in the connectivity of complex networks have been proved
to be important and have been used to explain the functionality,
robustness, stability, and structure of networks from Biology
\cite{maslov} and Sociology \cite{newman2} to Computer Science
\cite{vespignani2}.  A study of the correlation profile in a network
of protein-protein interactions revealed that links from hubs to
non-hub nodes are favored \cite{maslov}, a result with consequences
for the stability and the modularity in biological networks. In the
case of social networks, Newman has shown that most of them are
assortative (i.e. hub-hub correlations dominate the system)
\cite{newman2}, and Colizza {\it et al.} demonstrated the `rich-club'
phenomenon where all hubs tend to form a connected cluster
\cite{Colizza}. As a result, social networks are more difficult to
immunize and diseases can spread fast.

Recently, it was also shown that hub anticorrelations, i.e. the
tendency of the hubs not to be directly connected with each other,
give rise to fractal networks \cite{shm}, such as the undirected
(symmetrized) WWW, the protein homology network \cite{homology} and other biological networks. On the contrary, when
there is a large probability of direct hub connections the resulting
networks, such as the Internet, the cond-mat co-authorship and other
social networks, are non-fractals \cite{shm2}. In this category falls
also the random configuration model \cite{Molloy,ba}.

An important topological feature of complex networks is the degree
distribution $P(k)$, where $k$ is the number of links for a given
node.
Although the form of $P(k)$ has a direct influence on the network
properties, it cannot convey all the information for the network
structure. Thus, two networks can have the same distribution $P(k)$
but with completely different topologies, determined by the presence
of degree correlations.  This structure can be captured by the
probability $P(k_1,k_2)$ that two nodes of degree $k_1$ and $k_2$ are
connected to each other, and by quantities derived from  $P(k_1,k_2)$, such as the
Pearson coefficient $r$, the average degree of nearest neighbors
$k_{\rm nn}$, etc.

Despite their importance, a general theoretical framework to describe
and characterize degree correlations in scale-free networks is still
work in progress. For an attempt to describe correlations using a
master equation approach see \cite{redner}. Here, we find that the
degree correlations in the studied scale-free networks can be
characterized in terms of a correlation exponent $\epsilon$, which we
calculate using a renormalization approach.  This allows us to propose
a classification of a set of dissimilar networks, according to the
degree of correlations into a small number of different classes in a
``phase diagram''. For example, biological networks and the WWW are in
the strong anti-correlations part of the diagram, while social
networks and the Internet are clustered near the region of random
networks. We show how we can use these ideas to explore more network
properties, such as diffusion and robustness, which depend on the
degree of correlations in the network.

We start by recalling the renormalization of a network under a scale
transformation.  The renormalization procedure tiles a network
according to the box-covering algorithm \cite{sghm}, with the minimum
number of boxes where the maximum distance in any box is less than
$\ell_B$. Each box is subsequently replaced by a node, and links are
established between these new `super-nodes' if at least one node
included in a box was connected to any node of the other box. 
These boxes are treated as the nodes of the renormalized network.
Renormalization is a reliable method for determining how the network
behaves at different length scales.  Self-similarity is then obtained
if the network structure remains invariant under the renormalization.

Alternatively, we can retain multiple links between the boxes when we
renormalize a network \cite{maslov2}. As we show below, though,
this is not a strong effect,
mainly due to preservation of the self-similar structure under renormalization.
In contrast, during a random rewiring process degree correlations are destroyed,
but bias is introduced if multiple links are forbidden.

\begin{figure}
\centerline{\resizebox{9cm}{!} { \includegraphics{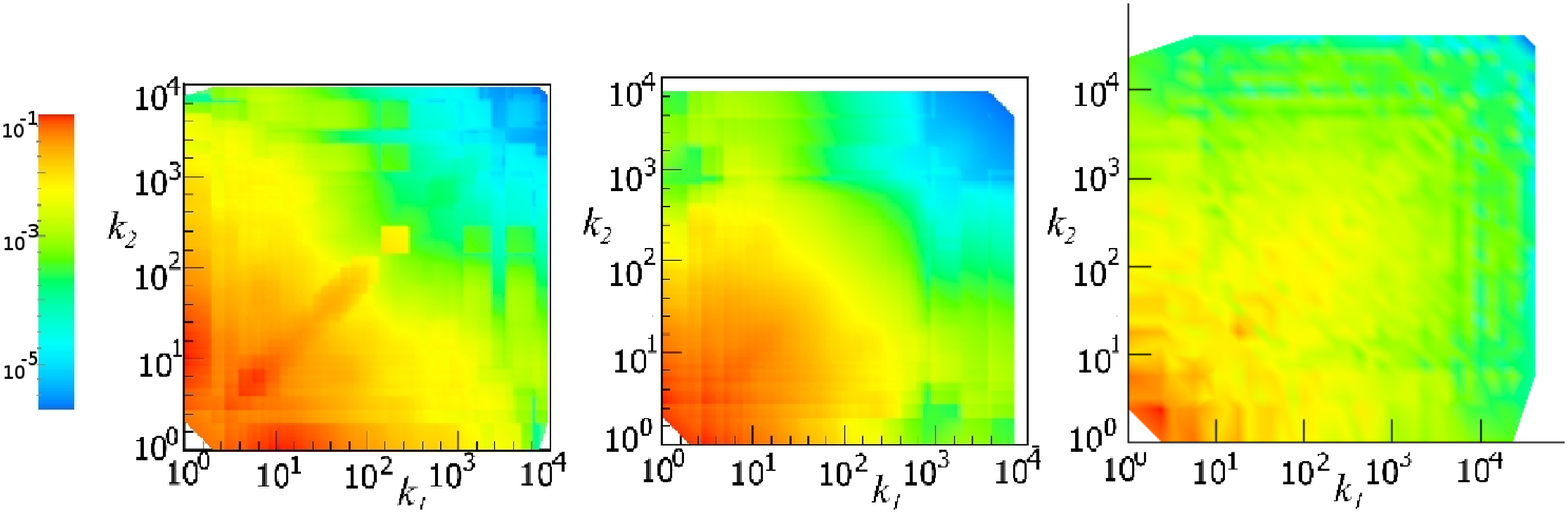}}}
\centerline{\resizebox{9cm}{!} {
\includegraphics{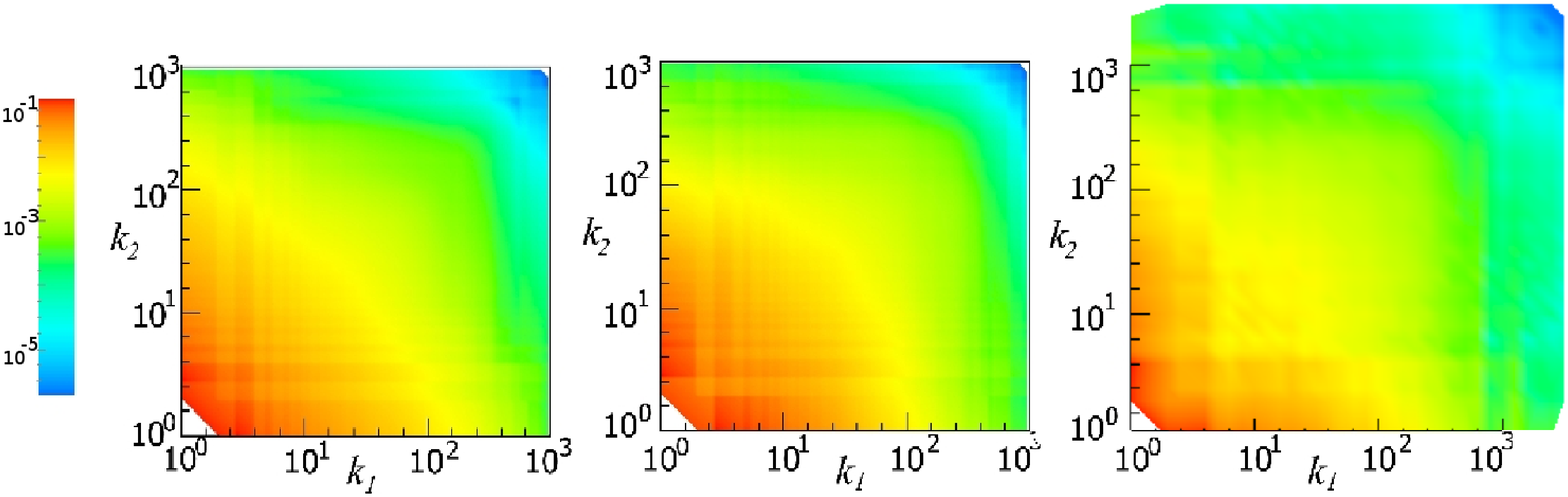}}}
 \caption {The joint degree distribution $P(k_1,k_2)$ of WWW (top row)
and Internet at the router level (bottom row) before renormalization (left), after
renormalization forbidding multiple links (center), and including multiple links (right).}
\label{pkk}\end{figure}

We use renormalization and scaling theory to determine the form
of $P(k_1,k_2)$.  Since the self-similarity of a scale-free network
requires the invariance of the degree distribution $P(k)$, a power-law
distribution of $P(k) \sim k^{-\gamma}$,
where $\gamma$ is the degree exponent, 
is the only form that can satisfy this
condition \cite{shm}. Taking this idea one step further, it is
interesting to clarify whether correlations between degrees, as
expressed by the joint degree distribution $P(k_1,k_2)$, also remain
invariant. In Fig. \ref{pkk} we present an example of this
distribution before and after renormalization for the WWW and the
Internet at the router level (similar results are derived for other
networks, as well). Allowing multiple links between boxes does not
significantly modify the result.
The statistical similarity of the corresponding plots
suggests the invariance of $P(k_1,k_2)$.  Accordingly, this suggests
that the $k_1$ and $k_2$ dependence can be separated and the behavior
of the tail of the joint degree distribution is:
\begin{equation}
\label{EQjoint}
P(k_1,k_2) \sim k_1^{-(\gamma-1)}k_2^{-\epsilon}\ (k_1 > k_2).
\end{equation}
The value of the first exponent $\gamma-1$ in Eq.~(\ref{EQjoint}) is 
obtained from the density
conservation law:
$\int P(k_1,k_2) dk_2 = k_1P(k_1) \sim k_1^{-(\gamma-1)}.$
Equation~(\ref{EQjoint}) is also consistent with the known result for
completely random networks
\begin{equation}
P(k_1,k_2) \sim k_1 P(k_1) k_2 P(k_2) \sim k_1^{1-\gamma} k_2^{1-\gamma} \,,
\label{EQrandom}
\end{equation}
i.e. the exponent $\epsilon$ for these networks is
$\epsilon_{\rm rand}=\gamma-1$, as expected from the symmetry in this case.
Eq. (\ref{EQjoint}) is also consistent with previous findings
in Ref.~\cite{maslov}. The probability distribution for the neighbor connectivity in
the yeast protein interaction network was there shown to behave differently for
low-degree nodes, where the $k$-dependence
was $k^{1-\gamma}$, and for large-degree nodes, with a $k^{-\gamma}$ dependence.
Using Eq.~(\ref{EQjoint}) we can see that integrating over $k_2$ for low-degree nodes
($k_1>k_2$) we retrieve the $k^{1-\gamma}$ dependence. For the case of hubs, where integration
is over $k_1$, the dependence on the degree is $k^{-\epsilon}$ and for the yeast protein interaction network
we have calculated that $\epsilon=\gamma$ (see Fig.~\ref{classification}). These results
are in agreement with the observed behavior in \cite{maslov}.

\begin{figure}
\centerline{ \resizebox{9cm}{!} { \includegraphics{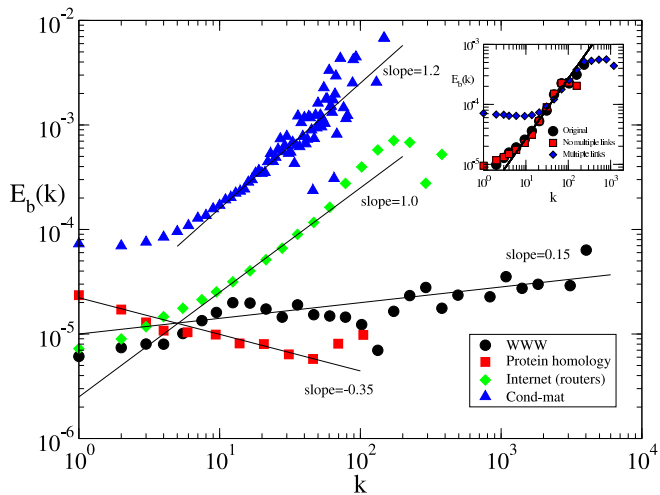}}}
\caption{
Plot of $E_b(k)$ versus $k$ for the WWW, protein homology, Internet at
the router level and cond-mat network.
Different topologies correspond to different scaling behavior with the
degree $k$. Inset: Plot of $E_b(k)$ versus $k$ for a) the Internet,
b) the renormalized Internet network without multiple links between
two nodes, and c) the renormalized Internet network allowing multiple links
between two nodes. The data have been vertically shifted in
order to show the invariance.
In b) and c) we use the MEMB method \cite{sghm} with $r_B=3$, and $b=3$.
}
\label{kk2}
\end{figure}

\begin{figure}
{\resizebox{9cm}{!} { \includegraphics{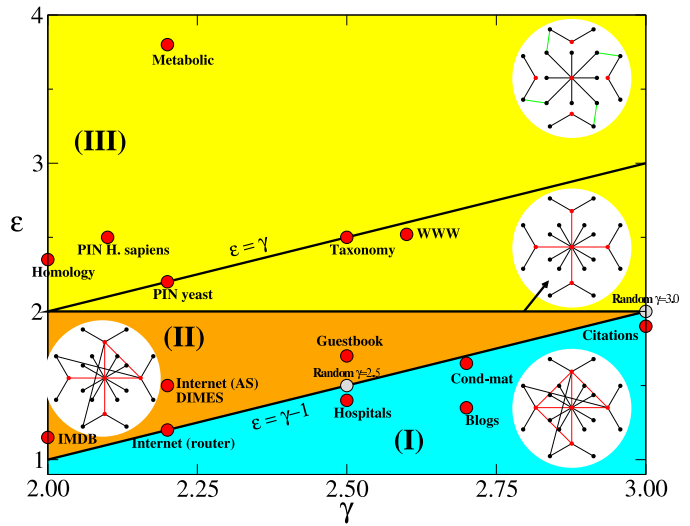}}}
\caption {Classification of scale-free networks \cite{databases}.  We use the
correlation exponent $\epsilon$ in order to quantify the degree of
correlations and the fractality of a network, as a function of
$\gamma$.
The line $\epsilon_{\rm rand}=\gamma-1$ corresponds to a completely
random network structure. The line $\epsilon=2$ separates fractal
($\epsilon>2$) from non-fractal networks ($\epsilon\leq 2$), while the
line $\epsilon=\gamma$ describes a fractal tree \cite{shm2}. The four
schematics illustrate networks where hub correlations are stronger
than in random networks (area I), weaker than random but non-fractal
(area II), non-fractal according to the minimal model of \cite{shm2}
($\epsilon=2$), and fractal (area III).    }
\label{classification}
\end{figure}

Next, we introduce a scale-invariant quantity $E_b(k)$ to simplify the
estimation of $\epsilon$, even for small networks.
We are motivated to introduce this quantity by asking whether a node is
significantly linked to more connected nodes, i.e. a node
considers another node as a `hub' if its degree is much larger than
its own. We define the ratio
\begin{equation}
E_b(k)\equiv \frac{\int_{bk}^{\infty} P(k|k')dk'}
{\int_{bk}^{\infty} P(k')dk'} \,,
\label{EQdefEbk}
\end{equation}
as the measure of a node's preference to connect to neighbors with
degree larger than $bk$ ($b$ is an arbitrary positive number, and
large $b$ corresponds to the identification of the hubs)
\cite{footnote}.  From the scaling of $E_b(k)$ with $k$, we are able to
obtain the exponent $\epsilon$ in a simpler way than using
$P(k_1,k_2)$, which presents more fluctuations than the average
quantity $E_b(k)$.  The conditional probability is $P(k|k') =
P(k,k') / \int P(k,k')dk = P(k,k') / k'^{1-\gamma} =
k^{-(\gamma-1)} k'^{-(1+\epsilon-\gamma)}$.
We find for a scale-free distribution:
\begin{equation}
E_b(k) \sim
k^{-(\epsilon-\gamma)} \,.
\label{cor}
\end{equation}
We have verified that the scaling of $E_b(k)$ remains invariant under
renormalization. The same scaling exponents are recovered for the
renormalized networks, even when multiple links are allowed between two boxes
(Fig.~\ref{kk2} inset). In the latter case, the renormalized nodes have in general
larger degrees, which means that deviations appear in nodes of smaller degree.
Additionally, $\epsilon$ was found to be independent of the value of $b$.
We notice that other quantities derived from
$P(k_1,k_2)$ may not be invariant under renormalization, such as $r$
or $k_{\rm nn}$, and therefore are not suitable to distinguish fractal
from non-fractal networks.

In Fig.~\ref{kk2} we present the behavior of $E_b(k)$ for the WWW,
protein homology, Internet (router level) and cond-mat authorship. The
existence of a scaling relation over a $k$ range, combined with the
invariance of this curve, support Eq.~(\ref{cor}) and the form used
for $P(k_1,k_2)$ in Eq.~(\ref{EQjoint}). The WWW and the protein
homology network have been shown to have a fractal topology. The slope
of $E_b(k)$ with $k$ is small or negative in these cases with values
of $\epsilon=2.5$ and $\epsilon=2.4$, respectively.  This behavior is
in contrast with the two non-fractal networks in the figure, i.e. the
Internet at the router level and the cond-mat co-authorship network,
where $E_b(k)$ increases almost linearly with increasing $k$. For
these networks we find that $\epsilon=1.2$ and $\epsilon=1.6$,
respectively.

In order to interpret the values of $\epsilon$ we now turn to the
renormalization scheme \cite{shm}.  After renormalization the number of
nodes $N$ in the network and the degree of a node $k$ scale with
$\ell_B$ as power laws with fractal exponent $d_B$ and degree exponent
$d_k$, respectively (we use a prime to describe quantities measured in
the renormalized network):
\begin{equation}
N \to N' \sim \ell_B^{-d_B} N \,\,\,,\,\,\, k \to k' \sim
\ell_B^{-d_k} k \,.
\label{EQdefN}
\end{equation}
If $d_B$ and $d_k$ are finite, the network is fractal. If $d_B\to
\infty$ and $d_k\to\infty$ (or equivalently the decay is exponential
or faster) the network is not fractal.

After tiling the network with boxes of diameter $\ell_B$, each
of these boxes have one unique local hub (i.e. the largest degree node
in the box).  Considering all possible pairs of boxes, we introduce
the probability ${\cal E} (\ell_B)$ that there exists a direct
connection between the two hubs of any two boxes.
We have shown (see e.g. Figs. 2e, 3d of
Ref.~\cite{shm2}) that the probability $\cal E$ scales with the
length $\ell_B$ as
\begin{equation}
\label{EQde_def}
{\cal E} (\ell_B) \sim \ell_B^{-d_e} \,.
\end{equation}
Below we  relate the exponent $\epsilon$ to the hub-hub
repulsion through the hub correlation exponent $d_e$, which is crucial
for fractality.

The conservation of links in the renormalized network leads to the expression
\begin{equation}
N P(k_1,k_2)dk_1dk_2 = \mathcal{E}(\ell_B)N' P'(k_1',k_2')dk_1'dk_2'
\,.
\label{EQstart}
\end{equation}
Using Eqs.~(\ref{EQjoint}), (\ref{EQdefN}), (\ref{EQde_def}), and
(\ref{EQstart}) we get the relation
$\ell_B^{d_B}\ell_B^{d_e}\ell_B^{(3-\gamma-\epsilon) d_k}=1$ which
finally leads to
\begin{equation}
\epsilon=2+d_e/d_k = 2 + (\gamma-1) \frac{d_e}{d_B} \,,
\label{EQepsilon}
\end{equation}
where we have substituted the value $\gamma = 1 + d_B/d_k$.  This
relation of $\epsilon$ with $d_e$ shows that correlations between the
hubs of the boxes determine the correlations for all degrees, in
accordance with the invariance under renormalization.

The direct determination of $\epsilon$ through the slope of $E_b(k)$
vs $k$ enables us to construct a `phase diagram' in the plane
$(\epsilon,\gamma)$, shown in Fig.~\ref{classification}.  This plot is
classifying the studied networks in classes according to their degree
of correlations, even though they correspond to dissimilar systems in
biology, sociology or technology. 

As shown in Eq.~(\ref{EQrandom}), the exponent for a random network
corresponds to the random line $\epsilon_{\rm rand}=\gamma-1$, which
is verified in the plot for different $\gamma$ values of the
configuration model.  In random network models, correlations arise
because links are selected for connecting with each other
equiprobably, so that the probability of two hubs being connected is
large \cite{catanzaro}.
Thus, networks that are close to the line $\epsilon_{\rm
rand}=\gamma-1$ exhibit hub-hub correlations.  The random line
separates the diagram in two main parts: (a) above the line where the
hub correlations tend to become weaker, and (b) below the line where
networks have even larger correlations (hubs are connected to each
other with even higher probability than the one corresponding to a
randomly created structure).

In the diagram, the social networks and the Internet at the router
level are clustered around the line $\epsilon_{\rm
rand}=\gamma-1$. This is an indication that there is a strongly
connected core of hubs in these systems, consistent with previous
studies.  The biological networks and the WWW, on the other side, are
far away from the random line. This implies that there is a richer
structure in these networks with hubs separated from each other. The
distance in the plot from $\epsilon_{\rm rand}$ quantifies how
different from randomness the network structure is, in terms of degree
correlations.

As $\epsilon$ increases from $\epsilon_{\rm rand}$, we expect that at
some point the networks will become fractal, due to increased hub-hub
correlations.  The point of emergent fractality is found through
Eq.~(\ref{EQepsilon}), where the borderline case of $d_B \to \infty$
yields $\epsilon=2$.  Indeed, we have verified via direct measurements
of $d_B$ that all the networks above the line $\epsilon=2$ in
Fig.~\ref{classification} are fractals.

Thus, starting from $\epsilon_{\rm rand}$ we can separate the phase
space into areas where the hub correlations are stronger than in
random models (area I) or weaker than that (areas II and III). The
weak correlation areas II and III are further divided by the line
$\epsilon=2$ which determines whether the anticorrelations are strong
enough to result in a fractal network (III) or not (II).

An immediate result from this diagram is the different
position of the Internet at the router level compared to the AS
level \cite{vazquez}. Although the degree distribution of these two networks is the
same ($\gamma=2.2$), the correlation exponent $\epsilon$ reveals that
there are more hub-hub connections at the router level, similar to the
case of a random network.  Contrary to that, the AS level exhibits a
structure with less correlations deviating from that of a simple
random model. This difference may hint on different design principles
or requirements at varying levels of the Internet.

\begin{figure}
{\resizebox{9cm}{!} { \includegraphics{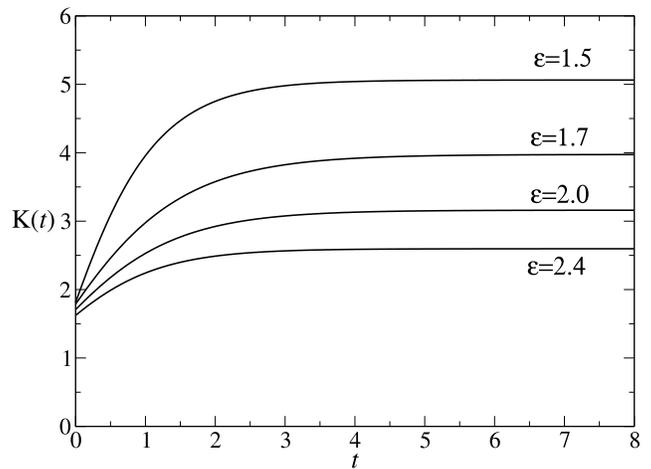}}}
\caption {Influence of correlations on diffusion, for scale-free
networks with $\gamma=2.75$.  The presented $\epsilon$ values
correspond to different areas in Fig.~3.  Decreasing hub-hub
correlations (top to bottom) leads to faster convergence towards
equilibrium, i.e. diffusion is accelerated.}
\label{fig4}
\end{figure}

The above approach can be directly applied to explore many interesting
properties, such as network robustness, synchronization, or diffusion
processes. Until now, theoretical studies have been limited to using
the uncorrelated version of $P(k'|k)\sim k' P(k')$. The introduction of
Eq.~(\ref{EQjoint}) enables us to substitute this form and generalize
the problem for networks with known correlation exponents. For
example, we can study the effect of correlations on diffusion by
starting with the master equation for the density of particles
$\rho(k,t)$ on nodes with degree $k$ at time $t$
\begin{equation}
\frac{d\rho(k,t)}{dt} = -\rho(k,t) + k \sum_{k'=k_{\rm min}}^\infty P(k'|k)
\frac{\rho(k',t)}{k'} \,,
\end{equation}
and substitute $P(k'|k)$ with a form derived through
Eq.~(\ref{EQjoint}).  The Laplace transform of the above equation
leads to a Fredholm integral equation of the second kind with
separable kernel that can be solved analytically. We define the
quantity $K(t)=\langle k^x (t) \rangle^{1/x}$, where $x=\gamma-1-\epsilon$,
and $K(t)$ serves as a measure of the diffusing particles preference to larger or smaller degrees $k$.
The analytical result for $\kappa_x(t)$, defined as $\kappa_x(t)=\langle k^x
\rangle(t)$, is $\kappa_x(t) = \kappa_\infty +
(\kappa_0-\kappa_\infty)e^{-ct}$, where
$\kappa_\infty=(\gamma-2)/(\epsilon-1)$ and
$\kappa_0=(\gamma-1)/\epsilon$ are constants depending on $\gamma$ and
$\epsilon$, while $c=(\epsilon-1)^2/((\gamma-2)(2\epsilon-\gamma))$.
The result for $K(t)$ is displayed in Fig.~\ref{fig4} for networks
with $\gamma=2.75$.  The exponential convergence to the asymptotic
steady-state configuration depends on the value of $c$, which
increases with the exponent $\epsilon$.  Networks that are close to
the random case $\epsilon_{\rm rand}=\gamma-1$ require longer times
for reaching the equilibrium state and the diffusing particles prefer
to occupy larger degree nodes. A larger $\epsilon$ value enhances
anti-correlations in the network and the particles move faster
occupying smaller degree nodes on the average.  We can infer, thus,
that stronger correlations tend to speed up the diffusion process.
The mechanism behind this behavior is as follows: when the hubs are
directly connected to each other the particles tend to remain
localized in the neighborhood around these hubs, so that it takes
longer for them to explore wider areas. On the contrary, when hub
anti-correlations are important the particles spend most of their time
in the intermediate areas which are formed by smaller degree nodes and
which connect indirectly the hubs to each other.

\begin{acknowledgments}
We acknowledge valuable discussions with Shlomo Havlin and support from NSF grants.
\end{acknowledgments}


\begin{thebibliography}{}
\bibitem{footnote} Notice that we could have defined $E_b(k)$ without
using the denominator in Eq.~(\ref{EQdefEbk}), i.e. $E_b(k)\equiv
\int_{bk}^{\infty} P(k|k')dk'$, in which case we get $E_b(k)\sim
k^{-(1-\epsilon)}$. However, due to the large fluctuations occurring at
the tails of the degree distribution in real-life networks, the
calculation of $\epsilon$ through Eq.~(\ref{EQdefEbk}) was proved to
be more robust.
\end{thebibliography}
\end{document}